\documentclass[12pt,preprint]{aastex}

\slugcomment{Computational Astrophysics Laboratory}

\shorttitle{Pulsar magnetospheric plasma distribution}
\shortauthors{Mc Donald, John, Shearer, Andrew}

\begin{document}

\title{Investigations of the magnetospheric plasma \\ 
distribution in the vicinity of a pulsar - I Basic formulation}

\author{John Mc Donald \& Andrew Shearer}
\affil{Computational Astrophysics Laboratory, Centre for Astronomy,\\
National University of Ireland, Galway, Ireland}

\begin{abstract}
The magnetospheric plasma distribution in the vicinity of a pulsar at various inclination angles is investigated using a new relativistic, parallel 3D Particle-In-Cell (PIC) code DYMPHNA3D. DYMPHNA3D uses a superposition of the electromagnetic fields associated with a rotating magnetised conducting sphere in a vacuum (\textit{the pulsar fields}) and the electromagnetic fields due to the presence of the magnetospheric plasma surrounding the pulsar (\textit{the plasma fields}), as the total fields. The plasma is moved self-consistently through the magnetosphere using the PIC methodology. Our initial simulation results are presented here. These show similar solutions to those obtained from previous numerical simulations, which show the fundamental plasma distribution in the vicinity of an aligned rotating neutron star to consist of two polar domes and an equatorial torus of trapped non-neutral plasma of opposite sign. The aligned case being the case in which the inclination angle, $\alpha$, between the magnetic dipole moment and the rotation axis of the star is zero. Furthermore, our code allows for off-axis simulations and we have found that this plasma distribution collapses into a `Quad-Lobe' charge-separated non-neutral magnetospheric plasma distribution in the case of an orthogonal rotator, i.e., the case in which the magnetic dipole moment is oriented at right angles to the rotation axis of the neutron star, with the plasma remaining trapped close to the stellar surface by the force-free ($\textbf{E}\cdot\textbf{B}=0$) surfaces. We find that if initialised with a Goldreich-Julian type distribution, the system is seen to collapse rapidly into these stable `Dome-Torus' structures. 
\end{abstract}


\section{Introduction}
The problem as to exactly how and where the observed emission from pulsars is generated has defied explanation for nearly four decades. There are many popular models, each with its own advantages and disadvantages, each with its dedicated supporters. These models differ primarily in terms of the location within the magnetosphere at which the observed emission is generated, i.e., the location of the particle acceleration region. The primary high-energy emission models at present are the polar cap model \citep{HM1998}, the outer gap model \citep{CHR1986}, the slot gap model \citep{HM2004,HARDINGETAL2008}, and the two-pole caustic model \citep{DR2003}. The polar cap model proposes that the particle acceleration region is located close to the stellar surface in the open field region over the pulsar's polar caps. The outer gap model assumes that the vacuum gaps within which the particle acceleration occurs form in the outer magnetosphere along null charge surfaces. The slot gap model contains elements of both of these models. This model is an extension of the original polar cap model, required to overcome some problems that the polar cap model faced in terms of difficulties in producing emission beams wide enough to correlate with the wide pulse profiles observed. By allowing the gap to extend upwards near the polar cap rim, asymptotically approaching the boundary between the open and closed magnetospheric regions, delimited by the last closed field lines, the slot gap model is capable of producing these wider pulse profiles. Last, but not least, is the two-pole caustic model which assumes that the vacuum gap, in which the particle acceleration occurs, is a thin gap that extends from each of the polar caps to the light cylinder and is confined to the surface of the last open magnetic field lines. 
\par
\cite{GJ1969} determined that a magnetised neutron star would be surrounded `everywhere' by a charge density given by,
\begin{equation}
\label{gjchargedensity}
\rho_{GJ}=\frac{1}{4\pi}\mathbf{\nabla}\mathbf\cdot\mathbf{E}=-\frac{1}{2\pi c}\frac{\mathbf{\Omega}\mathbf\cdot\mathbf{B}}{1-|\frac{(\mathbf{\Omega}\mathbf\times\mathbf{r})}{c}|},
\end{equation}
where $\rho_{GJ}$ is the Goldreich-Julian charge density and $\mathbf{\Omega}$ is the rotational frequency of the star. This was done for the aligned case which is a simplification of the more complex inclined rotator problem. They assumed that plasma on open field lines (i.e., the field lines which do not close within the light cylinder distance) can stream to infinity, while plasma on closed field lines corotates with the neutron star. They also assumed that a wind would be generated at the light cylinder, attributable to centrifugal forces due to the rigid corotation of this magnetospheric plasma with the star.
\par
This model advocates a totally filled magnetosphere (TFM), in which the plasma is pulled from the pulsar surface to fill the magnetosphere with a charge density given by Eq.~(\ref{gjchargedensity}). There are some critical problems with this model to include the fact that if the plasma has to follow magnetic field lines, then it is impossible for the magnetospheric region surrounding an aligned rotator to fill with plasma. Also, at large distance from the stellar surface, positive particles would have to travel along magnetic field lines which have their feet firmly rooted at the polar regions of the star, where only negative particles exist. The assertion that there needs to be acceleration of particles from the polar caps to replace particles lost at large distances from the pulsar is unfounded and these issues have been well addressed by \cite{MICHEL1982,MICHEL2004,MICHEL2005}. 
\par
In stark contrast to the scenario in which the pulsar is surrounded `everywhere' by the Goldreich-Julian charge density are the results of explicit computer simulations, first performed by \cite{KPM1985A,KPM1985B}. The results of such simulations demonstrate that it is possible to construct stable self-consistent solutions involving charge-separated non-neutral plasma distributions in the form of two polar domes of trapped charged particles separated by a vacuum gap from an equatorial torus of trapped charged particles of the opposite sign, for the case of the aligned rotator, $\alpha=0^{\circ}$.
\par
The basic premise behind these early simulations was to let the quantised surface charge on the neutron star follow the electric forces quasi-statically, using an iterative approach until the force-free condition ($\mathbf{E}\cdot\mathbf{B}=0$) is satisfied everywhere. This is an $O(N^{2})$ problem, as such it is computationally intensive and consequently, the number of particles had to be kept reasonably small by choosing a large charge quantisation level. In addition, in order to take advantage of the inherent symmetry involved in the system, the charge elements were taken to be rings.
\par
These simulations were later reworked by \cite{SMT2001}. In this case, the particles were taken to be `super-particles', with each `super-particle' representing a finite density of discrete particles and carrying a charge that is very large in comparison to the electronic charge. As before, the code moves the `super-particles' iteratively to their equilibrium positions, i.e., $\mathbf{E}\cdot\mathbf{B}=0$ positions based on the electromagnetic fields in the magnetosphere which are due to the neutron star and all of the other particles.
\par
As mentioned, these simulation methodologies are N-body in nature and as a result, are restrictive in terms of the number of particles that can be used due to computational overheads. The solution invoked to overcome this limitation is to use a Particle-In-Cell (PIC) approach, as was done by \cite{SA2002}. To perform their simulations, they used a modified version of the publicly available 3D electromagnetic relativistic PIC code `TRISTAN' \citep{BUNEMAN1993}. Their approach used a superposition of the vacuum rotator fields, i.e., the electromagnetic fields associated with a rotating conducting magnetised sphere in a vacuum \citep{DEUTSCH1955} and the plasma fields, i.e., the electromagnetic fields due to the presence of the magnetospheric plasma, as the total electromagnetic fields. Deutsch wrote down by inspection, the vacuum rotator field equations (providing little insight into their derivation) which require a significant amount of processing to convert them into functions of $r,\:\theta\:\&\:\phi$. An explicit derivation of these field equations was performed by \cite{MICHELLI1999}, Eq.~(\ref{DEUTSCH_EQNS}).
\begin{eqnarray}
\label{DEUTSCH_EQNS}
&&B_{r}=2B_{0}\frac{a^{3}}{r^{3}}\left(\cos\alpha\cos\theta+\sin\alpha\sin\theta\cos\phi_{s}\right) \nonumber \\
&&B_{\theta}=B_{0}\frac{a^{3}}{r^{3}}\left(\cos\alpha\sin\theta-\sin\alpha\cos\theta\cos\phi_{s}\right) \nonumber \\
&&B_{\phi}=B_{0}\frac{a^{3}}{r^{3}}\sin\alpha\sin\phi_{s} \nonumber \\
&&\lefteqn{E_{r}=\Omega{a}B_{0}\Biggl(\frac{2}{3}\cos\alpha\frac{a^{2}}{r^{2}}+\cos\alpha\frac{a^{4}}{r^{4}}\left(1-3\cos^{2}\theta\right)} \nonumber \\
&&\hspace{4em}-3\sin\alpha\frac{a^{4}}{r^{4}}\sin\theta\cos\theta\cos\phi_{s}\Biggr) \nonumber \\
&&\lefteqn{E_{\theta}=\Omega{a}B_{0}\Biggl[-2\cos\alpha\frac{a^{4}}{r^{4}}\sin\theta\cos\theta} \nonumber \\
&&\hspace{4em}+\sin\alpha\left(\frac{a^{4}}{r^{4}}\cos2\theta-\frac{a^{2}}{r^{2}}\right)\cos\phi_{s}\Biggr] \nonumber \\
&&E_{\phi}=\Omega{a}B_{0}\sin\alpha\left(\frac{a^{2}}{r^{2}}-\frac{a^{4}}{r^{4}}\right)\cos\theta\sin\phi_{s},
\end{eqnarray}
\vspace{1em}
\par
where $B_{0}$ is the magnetic field strength at the equator, r is the radial distance from the star, $\theta$ is the stellar latitude, $\phi$ is the stellar longitude ($\phi_{s}=\phi-\Omega t$), and $\alpha$ is the inclination angle between the magnetic dipole moment and the rotation axis of the neutron star.
\par
Our simulations also use `super-particles', each of which represents a finite density of discrete particles and carries a charge that is very large in comparison to the electronic charge. The aligned rotator scenario mentioned above is the case of maximal symmetry. It is a simplification of the significantly more complicated inclined rotator scenario, in which the magnetic dipole moment is oriented at an angle, $\alpha$, to the rotation axis, i.e., an orthogonal rotator being the case of minimal symmetry. It has been argued \citep{MICHEL2005} that focusing efforts on the aligned case is a futile endeavour as this case does not produce an active pulsar. Also, the physics of this scenario is so drastically different to that of the inclined/orthogonal case, that the aligned model is incapable of generating any real insight into the operation of pulsars. Instead it is contested that the inclination between the magnetic and rotational axes of the neutron star is an essential attribute of any (\textit{active}) pulsar model \citep{MICHELSMITH2001,MICHEL2005}. 
\par
These simulations have independently verified the original KPM simulation results and showed the magnetospheric charge distribution to consist of these two polar domes of trapped negative charge (\textit{electrons}), with an equatorial torus of trapped positive charge (\textit{protons, positrons or ions}), separated by a vacuum gap. It is worth noting that the fundamentals of this `Dome-Torus' magnetospheric plasma distribution have also been investigated and verified independently by \cite{RYLOV1976, RYLOV1989, SHIBATA1989, NEUKIRCH1993, ZACHARIADES1993, THIELHEIM1994}, and \cite{PETRIETAL2002} as indicated by \cite{MICHEL2004}.
\section{Simulation code}
The code currently under development is a 3D fully electromagnetic, parallel, scalable, and relativistic PIC code, which has been designed in an extremely modular fashion. The codes modular design will facilitate ease of application of the code to a wide variety of kinetic plasma phenomena with little modification.
\par
Due to its inherent simplicity, PIC methodology is widely used for the kinetic simulation of plasmas. Its implementation involves the imposition of a spatial mesh (\textit{a spatial discretisation}) on the region containing the plasma. A staggered mesh system known as the Yee lattice \citep{YEE1966} is commonly utilised for this purpose. Electromagnetic field quantities are defined on the spatial grid whereas particles can occupy arbitrary locations in continuum space. The attributes of the plasma, i.e., the charge density and current density (\textit{field sources}) are deposited onto the spatial mesh and thus, the plasma particles need only interact with the mesh, as opposed to each other as in the case of a Coulomb problem. This reduces a computationally infeasible $O(N^{2})$ problem to a computationally viable $O(N\log N)$ scale. Once the plasma attributes have been deposited onto the spatial mesh, Maxwell's equations are self-consistently solved on the mesh to determine the electromagnetic fields due to the presence of the plasma. These plasma fields are then superimposed with the vacuum rotator (\textit{pulsar}) fields, i.e., the Deutsch fields, to determine the total electromagnetic fields. The total fields, located on the mesh, are then interpolated back to the location of the particles, the Lorentz force calculated and the particles are moved by integration of Newton's laws of motion. This procedure is iterated (\textit{a temporal discretisation}) self-consistently moving the plasma through the computational domain in accordance with the total fields calculated above.
\subsection{Data output}
One of the major difficulties with modern large-scale simulation codes is the sheer volume of the output data produced in many cases. DYMPHNA3D is no exception and is capable of producing a very large amount of output data, depending on the scale of the simulation and the configuration of the desired output. With this in mind, throughout the development and construction of the code, we focused heavily on the data output for post-processing and visualisation of the results in order to take maximal advantage of the 3D nature of the simulation and to be able to efficiently process the output.
\par
The user has the option of outputting the required simulation data at the end of each iteration (or specified increment of iterations), in order to study the temporal evolution or just the final state at the end of the simulation run if that is all that is required. The choice of data to output includes the Cartesian components of the particle locations, velocities, the Lorentz force on the particles, the electric field, the magnetic field, and the current density. The particle energies, their gamma factors, $\mathbf{E}\cdot\mathbf{B}$ on the mesh, and the charge density on the grid as well as specific phase space data for the particles can also be output if required. All data can be output in either standard data files for arbitrary post-processing or in VTK (Visualisation Toolkit, http://www.vtk.org) file format.  
\section{Numerical model}
As an initial test of the code, the investigation of the magnetospheric plasma distribution in the vicinity of a pulsar was chosen. The PIC code has been imposed on the region surrounding the pulsar, having the star centred in the computational domain. For simplicity, we have assumed that the positive and negative charge carrying species are of equal mass, alleviating any mass disparity complications in the computational model. We have initialised a neutral plasma (surface charge) on the pulsar surface, i.e., positive and negative charge carrying `super-particle' pairs are initialised at the same location on the stellar surface. This is done in order to satisfy initial conditions requiring charge neutrality, such that Gauss' law, Eq.~(\ref{GUASS_LAW}), is satisfied initially and as long as the charge continuity equation, Eq.~(\ref{CONTINUITY_EQN}), is not violated, it will be satisfied at subsequent times during the simulation.
\begin{equation}
\label{GUASS_LAW}
\mathbf{\nabla}\mathbf\cdot\mathbf{E}=-\frac{1}{\epsilon_{0}}\rho
\end{equation}
\vspace{1em}
\begin{equation}
\label{CONTINUITY_EQN}
\mathbf{\nabla}\mathbf\cdot\mathbf{J}=-\frac{\partial\rho}{\partial t}
\end{equation}
Taking Maxwell's $\mathbf{\nabla}\mathbf\cdot\mathbf{B}\:=\:0$ equation as a second initial condition, leaves Maxwell's two curl equations, from which a set of update equations for the electromagnetic fields due to the presence of the magnetospheric plasma can be derived, Eqs.~\ref{EX_UPDATE}-\ref{BZ_UPDATE}.
\subsection{Simulations}
\begin{figure*}[t]
\figurenum{1}
\epsscale{1.0}
\plotone{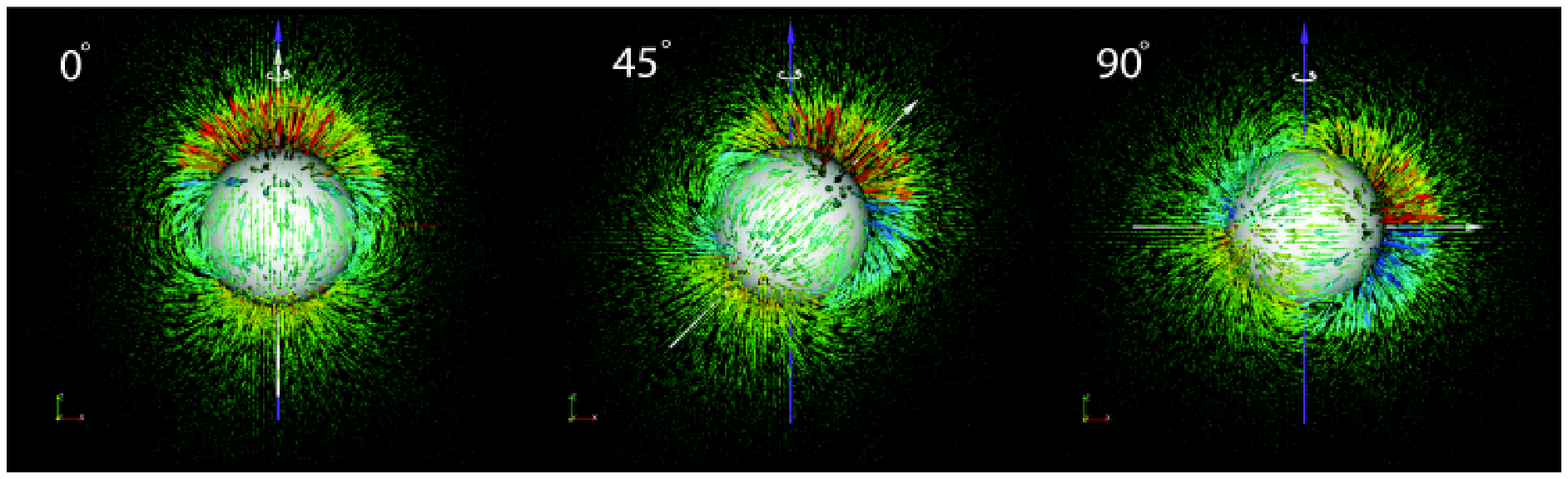}
\figcaption{\label{B_FIELD_EVOLUTION} 3D vector plots of the magnetic field structure in the vicinity of a pulsar for various angles of inclination, $\alpha$, from Paraview, ranging from the aligned case (\textit{leftmost panel}) through $45^{\circ}$ inclination (\textit{centre panel}) to the orthogonal case (\textit{rightmost panel}). The colour scheme tends from red (\textit{maximum magnitude vector facing upwards}) through orange, yellow, centred on green, through turquoise and light blue to dark blue (\textit{maximum magnitude vector facing downwards}). It is evident from these images that the magnetic field is seen to retain its dipolar form throughout the inclination from aligned to orthogonal rotator. Note that it is the length of the field vectors and their respective directions that gives the illusion of anti-symmetry in the images, the fields are symmetrical. We plot only a portion of the total number of field vectors, for the purpose of clarity, which also aids in the illusion of anti-symmetry. The purple axis denotes the stellar rotation axis while the white axis marks the magnetic dipole moment. Image taken looking in along the negative y-axis towards the origin.}
\end{figure*}

\begin{figure*}[t]
\figurenum{2}
\epsscale{1.0}
\plotone{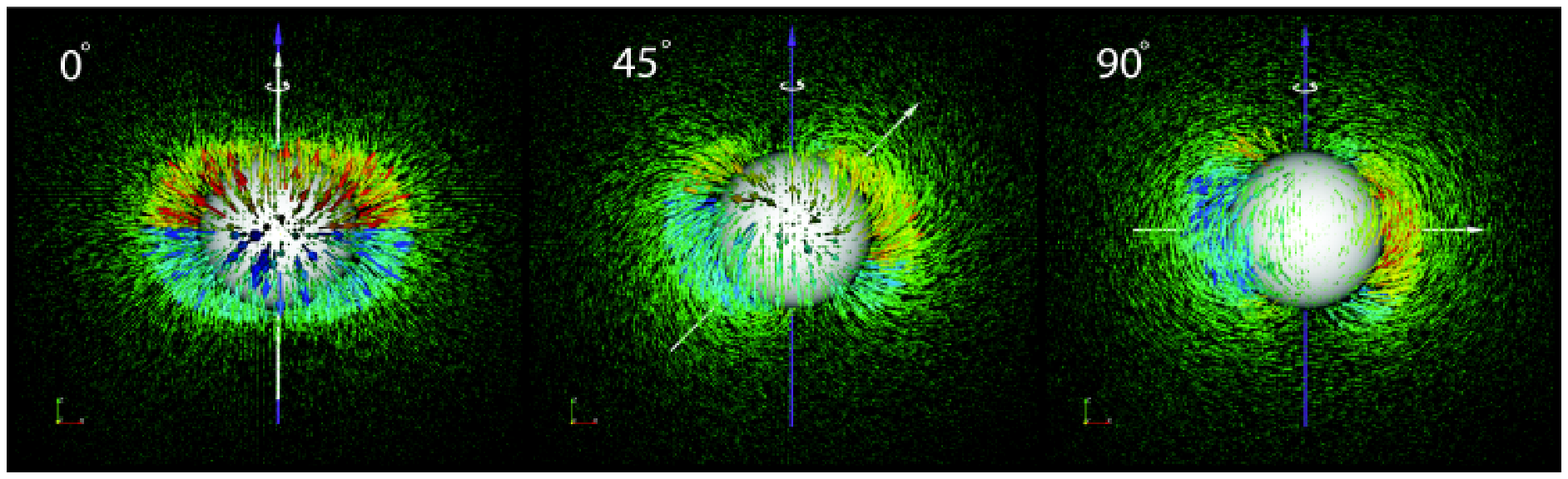}
\figcaption{\label{E_FIELD_EVOLUTION} 3D vector plots of the electric field structure in the vicinity of a pulsar for various angles of inclination, $\alpha$, from Paraview, ranging from the aligned case (\textit{leftmost panel}) through $45^{\circ}$ inclination (\textit{centre panel}) to the orthogonal case (\textit{rightmost panel}). The colour scheme tends from red (\textit{maximum magnitude vector facing upwards}) through orange, yellow, centred on green, through turquoise and light blue to dark blue (\textit{maximum magnitude vector facing downwards}). It is clear from the images that there is a significant evolution of the electric field structure from its initial quadrupolar nature during the inclination. We plot only a portion of the total number of field vectors, for the purpose of clarity. The purple axis denotes the stellar rotation axis while the white axis marks the magnetic dipole moment. Image taken looking in along the negative y-axis towards the origin.}
\end{figure*}

Simulations were run on the 32 processor BULL Novascale shared-memory machine at the ICHEC\footnotemark[1]\footnotetext[1]{Irish Centre for High End Computing, http://www.ichec.ie} facility. As discussed above, the total electromagnetic fields utilised in the simulations are a superposition of the plasma fields and the pulsar fields. The electromagnetic field structure in the vicinity of the pulsar has been visualised by means of 3D vector plots using Paraview (http://www.paraview.org). Figures~\ref{B_FIELD_EVOLUTION}\footnotemark[2] \& \ref{E_FIELD_EVOLUTION}\footnotemark[2]\footnotetext[2]{Note that it is the length of the field vectors and their respective directions that gives the illusion of anti-symmetry in the images, the fields are symmetrical. We plot only a portion of the total number of field vectors, for the purpose of clarity, which also aids in the illusion of anti-symmetry. The purple axis denotes the stellar rotation axis while the white axis marks the magnetic dipole moment.} illustrate the structure of the fields for various angles of inclination, $\alpha$, ranging from the aligned rotator scenario, $\alpha\:=\:0^{\circ}$ (\textit{left}), to the case of the orthogonal rotator, $\alpha\:=\:90^{\circ}$ (\textit{right}). The images in these two figures are taken looking in along the negative y-axis towards the origin. The dipolar nature of the magnetic field is overtly evident in the leftmost panel of Fig.~(\ref{B_FIELD_EVOLUTION}) and is seen to retain this dipolar form throughout the inclination from aligned rotator, $\alpha\:=\:0^{\circ}$, to orthogonal rotator, $\alpha\:=\:90^{\circ}$. 
\par
The case however, is very different for the electric field during inclination. In the leftmost panel of Fig.~(\ref{E_FIELD_EVOLUTION}), i.e., for the aligned rotator scenario, the structure of the \textbf{E} field can be observed to be primarily quadrupolar in nature, but its structure is seen to evolve significantly with inclination. From a visual inspection of the set of equations in Eq.~(\ref{DEUTSCH_EQNS}), it can clearly be seen that they are simply a superposition of the aligned ($\cos\alpha$ terms, symmetrical field structure, with no time-dependence) and orthogonal ($\sin\alpha$ terms, anti-symmetrical field structure, with time-dependence) cases. The simplicity of the aligned rotator scenario with respect to that of the orthogonal rotator is acutely evident from such an analysis of the field equations, illustrating the reasoning behind the historically predominant investigation of the former.
\par
In terms of our investigation of the plasma distribution in the vicinity of a pulsar, the initialised neutral plasma (surface charge) on the stellar surface is released into the magnetosphere and it self-consistently follows the ambient fields driven by the iterative PIC algorithm in accordance with the total determined fields, as discussed above. Once released, the particles are numerically free to travel anywhere within the computational domain (governed by the ambient electromagnetic fields), which includes reentering the star if they want to. With typical simulation runs containing of the order of 2M to 20M `super-particles', there is always a supply of charge available on the stellar surface for emission into the magnetosphere, wherever it is needed by the ambient fields. As mentioned above, the code can output the state of the simulation at each iteration (or increment of iterations), which allows us to study the temporal and spatial evolution of the magnetospheric plasma distribution.
\par
The simulation is terminated when a stable configuration for the non-neutral plasma is obtained. The condition for this is that the plasma has distributed itself, such that it eliminates the parallel component of the electric field along the magnetic field lines responsible for pulling the charges from the stellar surface and accelerating them out along the magnetic field lines. Thus, some form of trigger is required to terminate the simulation when this condition has been reached numerically. There are two relatively straight forward ways of testing for the presence of this stable configuration, firstly one can monitor the value of $\mathbf{E}\cdot\mathbf{B}$ inside the plasma itself, which should be zero everywhere within the plasma for it to have equilibriated. Secondly, the parallel component of the electric field along the magnetic field lines serves to pull charges from the stellar surface as mentioned above. Thus, if this parallel component of the electric field has been eliminated by the plasma redistributing itself to do so, then the existence of this condition will be marked by a cessation of particle emission from the stellar surface. This was essentially, the criterion that was utilised by \cite{SA2002}, to identify when this configuration had been achieved in their simulations and is also the trigger that we have employed in our model.
\par
As the simulation progresses, the code continually monitors the motion of particles released from the stellar surface for a cessation of emission into the magnetosphere. Once the code has determined that a cessation of emission of plasma has been achieved, i.e., particles are no longer being pulled from the surface of the star as a result of the parallel component of the electric field above the surface being eliminated, this triggers the termination of the simulation. We find that the stable equilibrium in the case of an aligned rotator, $\alpha\:=\:0^{\circ}$, is in the form of a charge-separated distribution consisting of two polar domes and an equatorial torus of trapped non-neutral plasma of opposite signs.
\subsection{Force-Free surfaces - Trapping regions}
\begin{figure*}[t]
\figurenum{3}
\epsscale{1.0}
\plotone{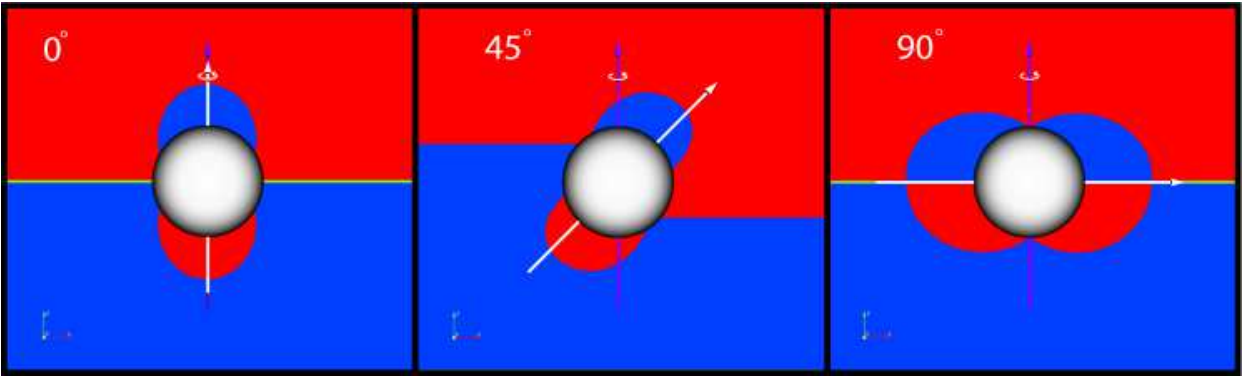}
\figcaption{\label{E_DOT_B_EVOLUTION} Plots of $\textbf{E}\cdot\textbf{B}$ in the vicinity of a pulsar for various inclination angles, $\alpha$, from Paraview, ranging from the aligned case (\textit{leftmost panel}) through $45^{\circ}$ inclination (\textit{centre panel}) to the orthogonal case (\textit{rightmost panel}). Red indicates that $\textbf{E}\cdot\textbf{B}\:>\:0$, blue indicates that $\textbf{E}\cdot\textbf{B}\:<\:0$ and green indicates $\textbf{E}\cdot\textbf{B}\:=\:0$. Images are taken in a slice through the pulsar origin in the x-z plane. The shape of the trapping regions is evident in the locus of the $\textbf{E}\cdot\textbf{B}\:=\:0$ surfaces in these plots. The purple axis denotes the stellar rotation axis while the white axis marks the magnetic dipole moment. It should be noted that this image depicts the vacuum rotator fields only, i.e., in the absence of plasma surrounding the star.}
\end{figure*}

The fundamental reason that we see these stable charge-separated non-neutral plasma structures is clearly evident when we look at $\mathbf{E}\cdot\mathbf{B}$ in the vicinity of the neutron star, taking a slice through the pulsar origin in the x-z plane for clarity. Plots of $\mathbf{E}\cdot\mathbf{B}$ in the vicinity of the stellar surface for various inclination angles, $\alpha$, have been included in Fig.~(\ref{E_DOT_B_EVOLUTION}). It should be noted that the images of $\mathbf{E}\cdot\mathbf{B}$ in Fig.~(\ref{E_DOT_B_EVOLUTION}) depict the vacuum rotator fields only, i.e., in the absence of plasma surrounding the star. In these images, red indicates that $\mathbf{E}\cdot\mathbf{B}\:>\:0$, blue indicates that $\mathbf{E}\cdot\mathbf{B}\:<\:0$ and green indicates $\mathbf{E}\cdot\mathbf{B}\:=\:0$. The geometry of these force-free surfaces is in good agreement with those illustrated in \cite{MICHELLI1999}. It can be seen in the aligned case, that the polar domes are a reflection of these force-free trapping surfaces over the caps of the pulsar with negative charge being extracted from the poles, driven by the quadrupolar electric field. In the case of the equatorial torus, there is a force-free surface along the equatorial plane, on which the particles would tend to gather if it were not for their inter-repulsion which pushes them out to form the observed torus constrained outwardly by the tightly curved magnetic field lines in this region. 
\par
In the case of the orthogonal rotator, depicted in the rightmost panel of Fig.~(\ref{E_DOT_B_EVOLUTION}), the `Quad-Lobe' structure is clearly evident in the locus of these force-free trapping regions in the vicinity of the stellar surface.
\subsection{Plasma distribution - Simulation results \& Discussion}
\begin{figure*}[t]
\figurenum{4}
\epsscale{1.0}
\plotone{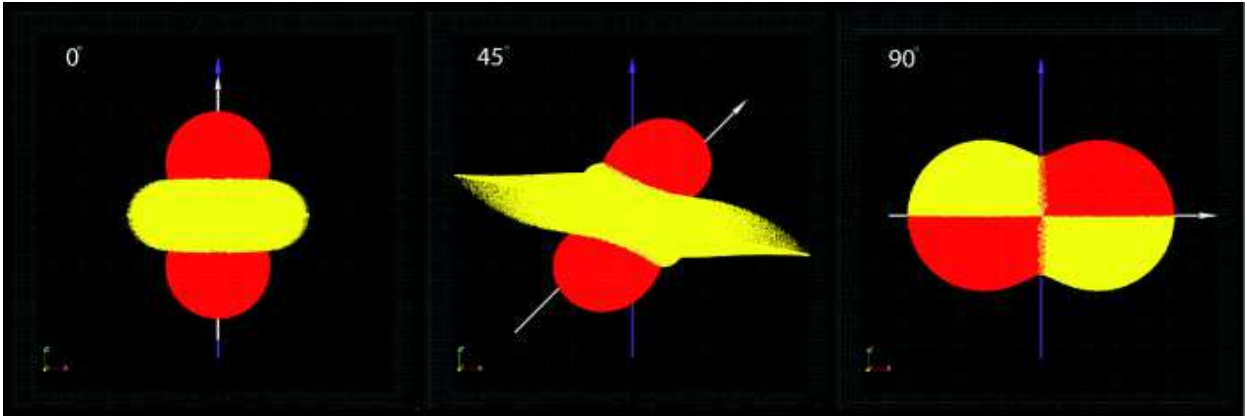}
\figcaption{\label{FINAL_PARTICLE_LOCATIONS} Plots of final particle locations in the vicinity of a pulsar for various inclination angles, $\alpha$, from Paraview, ranging from the aligned case (\textit{leftmost panel}) through $45^{\circ}$ inclination (\textit{centre panel}) to the orthogonal case (\textit{rightmost panel}). Red indicates the negative charge carrying species and yellow indicates the positive charge carrying species. The `Dome-Torus' structure can be seen in the aligned case in the leftmost panel and can be seen to collapse into this `quad-lode'-like structure in the orthogonal case in the rightmost panel. The purple axis denotes the stellar rotation axis while the white axis marks the magnetic dipole moment. Image taken looking in along the negative y-axis towards the origin.}
\end{figure*}

An illustration of the final particle locations predicted from our simulations has been included in Fig.~(\ref{FINAL_PARTICLE_LOCATIONS}), for the same angles of inclination included for the previous images, $\alpha\:=\:0^{\circ},\:45^{\circ}$ and $90^{\circ}$. In the case of the aligned rotator, the `Dome-Torus' charge-separated plasma distribution has been found to be the stable equilibrium condition, in agreement with the results of the previous numerical simulations performed. 
\par
As discussed above, this `Dome-Torus' distribution is not seen to persist with rotation of the magnetic dipole moment with respect to the stellar rotation axis. In the case of inclination to the orthogonal rotator scenario, i.e., $\alpha\:=\:90^{\circ}$, the `Dome-Torus' structure is observed to collapse into a `Quad-Lobe' type distribution as illustrated in the rightmost panel of Fig.~(\ref{FINAL_PARTICLE_LOCATIONS}). This `Quad-Lobe' structure bears a stark resemblance to the plasma distribution predicted by \cite{PARISH1974} (charge density given by Eq.~(\ref{PARISH_EQ})), with the exception that instead of being allowed to propagate out to a significant percentage of the light cylinder distance as was predicted in that early theoretical work, the plasma remains bound much closer to the stellar surface by these force-free trapping surfaces. 
\begin{equation}
\label{PARISH_EQ}
\rho=\frac{6\Omega B_{0}}{\mu_{0}c^{2}}\frac{a^{3}}{r^{3}}\sin\theta\cos\theta\left[1-\frac{\Omega^{2}r^{2}}{6c^{2}}(1-9\sin^{2}\theta)\right]
\end{equation}
\par
In the intermediate case, in which the inclination angle $\alpha=45^{\circ}$, the negative charge carrying species is seen to occupy the lopsided dome evident in the locus of the polar force-free surfaces at this inclination in Fig.~(\ref{E_DOT_B_EVOLUTION}). The positive charge carrying species is seen to propagate to a greater radial distance along the vertically displaced $\textbf{E}\cdot\textbf{B}=0$ surfaces that lie along the equatorial plane in the aligned case and are found to be stable in this configuration. 
\par
We have also found that if initialised with the Goldreich-Julian TFM state, the system is unstable and collapses into these stable trapped non-neutral plasma structures as was first demonstrated numerically by \cite{SMT2001}. 
\par
The fundamental problem with this scenario, in which the plasma is trapped in these non-neutral structures close to the stellar surface, is that observational evidence clearly illustrates that in practise, the plasma is not confined to these trapping zones. Images of the Crab nebula unambiguously depict a large-scale equatorial toroidal outflow and polar jets reminiscent to the topology of the `Dome-Torus' structure shown here. \cite{MICHEL2005} theorises that wave pickup of the charged particles by the large amplitude electromagnetic waves generated by the orthogonal component of the magnetic dipole field could be the resolution to this particular issue. 
\section{Conclusions}
The initial test application of this newly developed code has yielded some interesting results. We have found from our investigations into the plasma distribution in the vicinity of a rapidly rotating neutron star, that stable solutions involving charge-separated non-neutral plasma structures are not only viable, but are the natural equilibrium states for these systems. Our results agree well with the results of prior numerical simulations based on the aligned rotator scenario and we have extended these to look at the `off-axis' cases as well.
\par
The general topology of the Crab nebula, as observed in images from the Chandra X-ray observatory, is in some fashion reminiscent of the `Dome-Torus' type plasma distribution, evident in the observed equatorial toroidal outflow and polar jets. If the observed equatorial toroidal outflow and polar jets are in some way a reflection of this kind of plasma structure in the vicinity of the pulsar, then given that this distribution is seen to collapse into a `Quad-Lobe' type structure in the case of the orthogonal rotator, this may suggest that a nebula involving an orthogonal rotator or near orthogonal rotator, would not posses this observable macroscopic structure. This would also indicate that the inclination of the Crab pulsar should be appreciably lower than than that of an orthogonal rotator, in order that it still possess this macroscopic quasi-`Dome-Torus' structure. Chandra observations of the young Crab-like pulsar J0205+6449 in the SNR 3C 58 also clearly show a macroscopic torus structure (seen edge-on) and some jet-like structure. The inclination angle is estimated to be of the order of $70^{\circ}$, \cite{SLANEETAL2002,SLANEETAL2004}, slightly in excess of typical Crab estimates of around $65^{\circ}$, \cite{ZHANGCHENG2002}.  
\par
Given the success of this fledgling code in its initial application and its inherent modularity, these pave the way for application to a plethora of interesting astrophysical kinetic plasma phenomena. Future plans for the code involve continuing the initial investigations presented in this paper in greater detail to include probing the possibility and effects of the diocotron instability, \citep{SA2002,PETRIETAL2002,PETRIETAL2003}. Also planned as part of this work is the investigation of possible mechanisms required to perturb the observed stable distribution in order to allow the plasma to propagate beyond these trapping regions to form the observed equatorial outflow and polar jets. The nature of the developed code will allow us to numerically investigate the pulsar magnetospheric plasma distribution in unprecedented detail both spatially and temporally. Investigations of possible pulsar GRP mechanisms, necessarily over very short spatial scales are planned as well as simulations of a mechanism capable of generating broadband persistent electron cyclotron maser emission from Brown Dwarfs. The developed code is at its most basic, a `first-principles' approach to modelling plasmas and has been implemented in an extremely modular fashion facilitating the possibility of inclusion a wide variety kinetic phenomena into the computational model. 
\acknowledgments
The authors wish to thank the HEA Cosmogrid project for financial support and we acknowledge the SFI/HEA Irish Centre for High-End Computing (ICHEC) for the provision of computational facilities and support. F. C. Michel is thanked for his comments on some preliminary results for this paper. We would also like to thank the anonymous referee for comments which improved an earlier version of this paper.

\appendix

\section{Plasma fields update equations}
\textbf{The update equation for $E_{x}$:}
\noindent
\begin{eqnarray}
\label{EX_UPDATE}
\lefteqn{{{^{n+1/2}}{E_{x}}_{\left(i+1/2,j,k\right)}}={^{n-1/2}}{E_{x}}_{\left(i+1/2,j,k\right)}\:+\:} \nonumber \\
\nonumber \\
& & {c^{2}\Delta t}\Biggl(\frac{{^{n}{B_{z}}}_{\left(i+1/2,j+1/2,k\right)}\:-\:{^{n}{B_{z}}}_{\left(i+1/2,j-1/2,k\right)}}{\Delta y}\:-\: \nonumber \\
\nonumber \\
& & \frac{{^{n}{B_{y}}}_{\left(i+1/2,j,k+1/2\right)}\:-\:^{n}{B_{y}}_{\left(i+1/2,j,k-1/2\right)}}{\Delta z}\:-\:\mu_{0}\cdot^{n}{J_{x}}_{\left(i+1/2,j,k\right)}\Biggr)
\end{eqnarray}
\textbf{The update equation for $E_{y}$:}
\begin{eqnarray}
\label{EY_UPDATE}
\lefteqn{{{^{n+1/2}}{E_{y}}_{\left(i,j+1/2,k\right)}}={^{n-1/2}}{E_{y}}_{\left(i,j+1/2,k\right)}\:+\:} \nonumber \\
\nonumber \\
& & {c^{2}\Delta t}\Biggl(\frac{{^{n}{B_{x}}}_{\left(i,j+1/2,k+1/2\right)}\:-\:{^{n}{B_{x}}}_{\left(i,j+1/2,k-1/2\right)}}{\Delta z}\:-\: \nonumber \\
\nonumber \\
& & \frac{{^{n}{B_{z}}}_{\left(i+1/2,j+1/2,k\right)}\:-\:^{n}{B_{z}}_{\left(i-1/2,j+1/2,k\right)}}{\Delta x}\:-\:\mu_{0}\cdot^{n}{J_{y}}_{\left(i,j+1/2,k\right)}\Biggr)
\end{eqnarray}
\textbf{The update equation for $E_{z}$:}
\begin{eqnarray}
\label{EZ_UPDATE}
\lefteqn{{{^{n+1/2}}{E_{z}}_{\left(i,j,k+1/2\right)}}={^{n-1/2}}{E_{z}}_{\left(i,j,k+1/2\right)}\:+\:} \nonumber \\
\nonumber \\
& & {c^{2}\Delta t}\Biggl(\frac{{^{n}{B_{y}}}_{\left(i+1/2,j,k+1/2\right)}\:-\:{^{n}{B_{y}}}_{\left(i-1/2,j,k+1/2\right)}}{\Delta x}\:-\: \nonumber \\
\nonumber \\
& & \frac{{^{n}{B_{x}}}_{\left(i,j+1/2,k+1/2\right)}\:-\:^{n}{B_{x}}_{\left(i,j-1/2,k+1/2\right)}}{\Delta y}\:-\:\mu_{0}\cdot^{n}{J_{z}}_{\left(i,j,k+1/2\right)}\Biggr)
\end{eqnarray}
\textbf{The update equation for $B_{x}$:}
\begin{eqnarray}
\label{BX_UPDATE}
\lefteqn{{{^{n+1}}{B_{x}}_{\left(i,j+1/2,k+1/2\right)}}={^{n}}{B_{x}}_{\left(i,j+1/2,k+1/2\right)}\:+\:} \nonumber \\
\nonumber \\
& & \Delta t\Biggl(\frac{{^{n+1/2}{E_{y}}}_{\left(i,j+1/2,k+1\right)}\:-\:{^{n+1/2}{E_{y}}}_{\left(i,j+1/2,k\right)}}{\Delta z}\:-\: \nonumber \\
& & \frac{{^{n+1/2}{E_{z}}}_{\left(i,j+1,k+1/2\right)}\:-\:^{n+1/2}{E_{z}}_{\left(i,j,k+1/2\right)}}{\Delta y}\Biggr)
\end{eqnarray}
\textbf{The update equation for $B_{y}$:}
\begin{eqnarray}
\label{BY_UPDATE}
\lefteqn{{{^{n+1}{B_{y}}}_{\left(i+1/2,j,k+1/2\right)}}={^{n}{B_{y}}}_{\left(i+1/2,j,k+1/2\right)}\:+\:} \nonumber \\
\nonumber \\ 
& & \Delta t\Biggl(\frac{{^{n+1/2}{E_{z}}}_{\left(i+1,j,k+1/2\right)}\:-\:{^{n+1/2}{E_{z}}}_{\left(i,j,k+1/2\right)}}{\Delta x}\:-\: \nonumber \\
& & \frac{{^{n+1/2}{E_{x}}}_{\left(i+1/2,j,k+1\right)}\:-\:{^{n+1/2}{E_{x}}}_{\left(i+1/2,j,k\right)}}{\Delta z}\Biggr)
\end{eqnarray}
\textbf{The update equation for $B_{z}$:}
\begin{eqnarray}
\label{BZ_UPDATE}
\lefteqn{{{^{n+1}{B_{z}}}_{\left(i+1/2,j+1/2,k\right)}}={^{n}{B_{z}}}_{\left(i+1/2,j+1/2,k\right)}\:+\:} \nonumber \\
\nonumber \\
& & \Delta t\Biggl(\frac{{^{n+1/2}{E_{x}}}_{\left(i+1/2,j+1,k\right)}\:-\:{^{n+1/2}{E_{x}}}_{\left(i+1/2,j,k\right)}}{\Delta y}\:-\: \nonumber \\
& & \frac{{^{n+1/2}{E_{y}}}_{\left(i+1,j+1/2,k\right)}\:-\:{^{n+1/2}{E_{y}}}_{\left(i,j+1/2,k\right)}}{\Delta x}\Biggr)
\end{eqnarray}

\clearpage

\clearpage

\clearpage

\clearpage

\end{document}